\documentclass[letterpaper,10 pt]{IEEEtran}
\usepackage{microtype}
\usepackage{cite}
\usepackage{graphicx}
\usepackage{subcaption, subcaption}
\usepackage[para]{footmisc}
\usepackage{booktabs} %
\usepackage{comment}
\usepackage{makecell}
\usepackage{boldline}
\usepackage{amsfonts,amsmath,amssymb, amsthm} 
\usepackage{mathtools}
\usepackage{multirow}
\usepackage{microtype}
\usepackage{xspace}
\usepackage[ruled, vlined]{algorithm2e}
\usepackage{siunitx}
\DeclareSIUnit{\params}{\relax}
\usepackage{balance}
\newcommand{\pluseq}{\mathrel{{+}{=}}}
\newcommand{\minuseq}{\mathrel{{-}{=}}}
\usepackage[table]{xcolor}%
\newcommand{\PreserveBackslash}[1]{\let\temp=\\#1\let\\=\temp}
\newcolumntype{C}[1]{>{\PreserveBackslash\centering}p{#1}}

\usepackage{hyperref}
\usepackage{array}
\usepackage{tikz}
\newcolumntype{H}{>{\setbox0=\hbox\bgroup}c<{\egroup}@{}}

\newcommand{\SoundStream}{\emph{SoundStream}\xspace}
\newcommand{\waveform}{x}
\newcommand{\waveformhat}{\hat{x}}
\newcommand{\dims}{D}
\newcommand{\numsamples}{T}
\newcommand{\numquantizers}{N_q}
\newcommand{\numselectedquantizers}{n_q}

\newcommand{\numframes}{S}
\newcommand{\codebooksize}{N}
\newcommand{\samplerate}{f_s}
\newcommand{\bitrate}{R}
\newcommand{\rateperembedding}{r}
\newcommand{\ratio}{M}
\newcommand{\numblocks}{B}
\newcommand{\enc}{\text{enc}}
\newcommand{\dec}{\text{dec}}
\newcommand{\baseconvdepth}{C}
\newcommand{\quant}{Q}
\newcommand{\Loss}{\mathcal{L}}
\newcommand{\Exp}{E}
\newcommand{\G}{\mathcal{G}}  %
\newcommand{\D}{\mathcal{D}}  %
\newcommand{\numdiscr}{K}
\newcommand{\numdiscrout}{T_k}
\newcommand{\numlayers}{L}
\newcommand{\SPEC}{\mathcal{S}}
\newcommand{\visqol}{ViSQOL\xspace}
\newcommand{\targetsamples}{\texttt{targets}}
\newcommand{\inputsamples}{\texttt{inputs}}
\newcommand{\denoise}{\texttt{denoise}}
\newcommand{\true}{\texttt{true}}
\newcommand{\false}{\texttt{false}}

\newcommand{\windowlen}{W}
\newcommand{\hoplen}{H}
\newcommand{\numbins}{F}

\newcommand{\filmmultcoeff}{\gamma_{n,c}}
\newcommand{\filmaddcoeff}{\beta_{n,c}}
\newcommand{\activation}{a_{n,c}}
\newcommand{\activationtrans}{\widetilde{a}_{n,c}}

\newcommand{\confintsym}[2]{#1\scriptsize{ $\pm$ #2}}

\title{SoundStream: An End-to-End Neural Audio Codec}

\author{Neil Zeghidour, Alejandro Luebs, Ahmed Omran, Jan Skoglund, Marco Tagliasacchi}

\begin{document}

\maketitle

\begin{abstract}
We present \SoundStream, a novel neural audio codec that can efficiently compress speech, music and general audio at bitrates normally targeted by speech-tailored codecs. \SoundStream relies on a model architecture composed by a fully convolutional encoder/decoder network and a residual vector quantizer, which are trained jointly end-to-end. Training leverages recent advances in text-to-speech and speech enhancement, which combine adversarial and reconstruction losses to allow the generation of high-quality audio content from quantized embeddings. By training with structured dropout applied to quantizer layers, a single model can operate across variable bitrates from 3\,kbps to 18\,kbps, with a negligible quality loss when compared with models trained at fixed bitrates. In addition, the model is amenable to a low latency implementation, which supports streamable inference and runs in real time on a smartphone CPU. In subjective evaluations using audio at 24\,kHz sampling rate, \SoundStream at 3\,kbps outperforms Opus at 12\,kbps and approaches EVS at 9.6\,kbps. Moreover, we are able to perform joint compression and enhancement either at the encoder or at the decoder side with no additional latency, which we demonstrate through background noise suppression for speech.
\end{abstract}

\section{Introduction}
Audio codecs can be partitioned into two broad categories: waveform codecs and parametric codecs. Waveform codecs aim at producing at the decoder side a faithful reconstruction of the input audio samples. In most cases, these codecs rely on transform coding techniques: a (usually invertible) transform is used to map an input time-domain waveform to the time-frequency domain. Then, transform coefficients are quantized and entropy coded. At the decoder side the transform is inverted to reconstruct a time-domain waveform. Often the bit allocation at the encoder is driven by a perceptual model, which determines the quantization process. Generally, waveform codecs make little or no assumptions about the type of audio content and can thus operate on general audio. As a consequence of this, they produce very high-quality audio at medium-to-high bitrates, but they tend to introduce coding artifacts when operating at low bitrates. Parametric codecs aim at overcoming this problem by making specific assumptions about the source audio to be encoded (in most cases, speech) and introducing strong priors in the form of a parametric model that describes the audio synthesis process. The encoder estimates the parameters of the model, which are then quantized. The decoder generates a time-domain waveform using a synthesis model driven by quantized parameters. Unlike waveform codecs, the goal is not to obtain a faithful reconstruction on a sample-by-sample basis, but rather to generate audio that is perceptually similar to the original. 

Traditional waveform and parametric codecs rely on signal processing pipelines and carefully engineered design choices, which exploit in-domain knowledge on psycho-acoustics and speech synthesis to improve coding efficiency. More recently, machine learning models have been successfully applied in the field of audio compression, demonstrating the additional value brought by data-driven solutions. For example, it is possible to apply them as a post-processing step to improve the quality of existing codecs. This can be accomplished either via audio superresolution, i.e., extending the frequency bandwidth~\cite{li2021seanet}, via audio denoising, i.e., removing lossy coding artifacts~\cite{biswas2020aac}, or via packet loss concealment~\cite{stimberg2020}.

\begin{figure}[t]
    \centering
        \includegraphics[width=0.85\columnwidth]{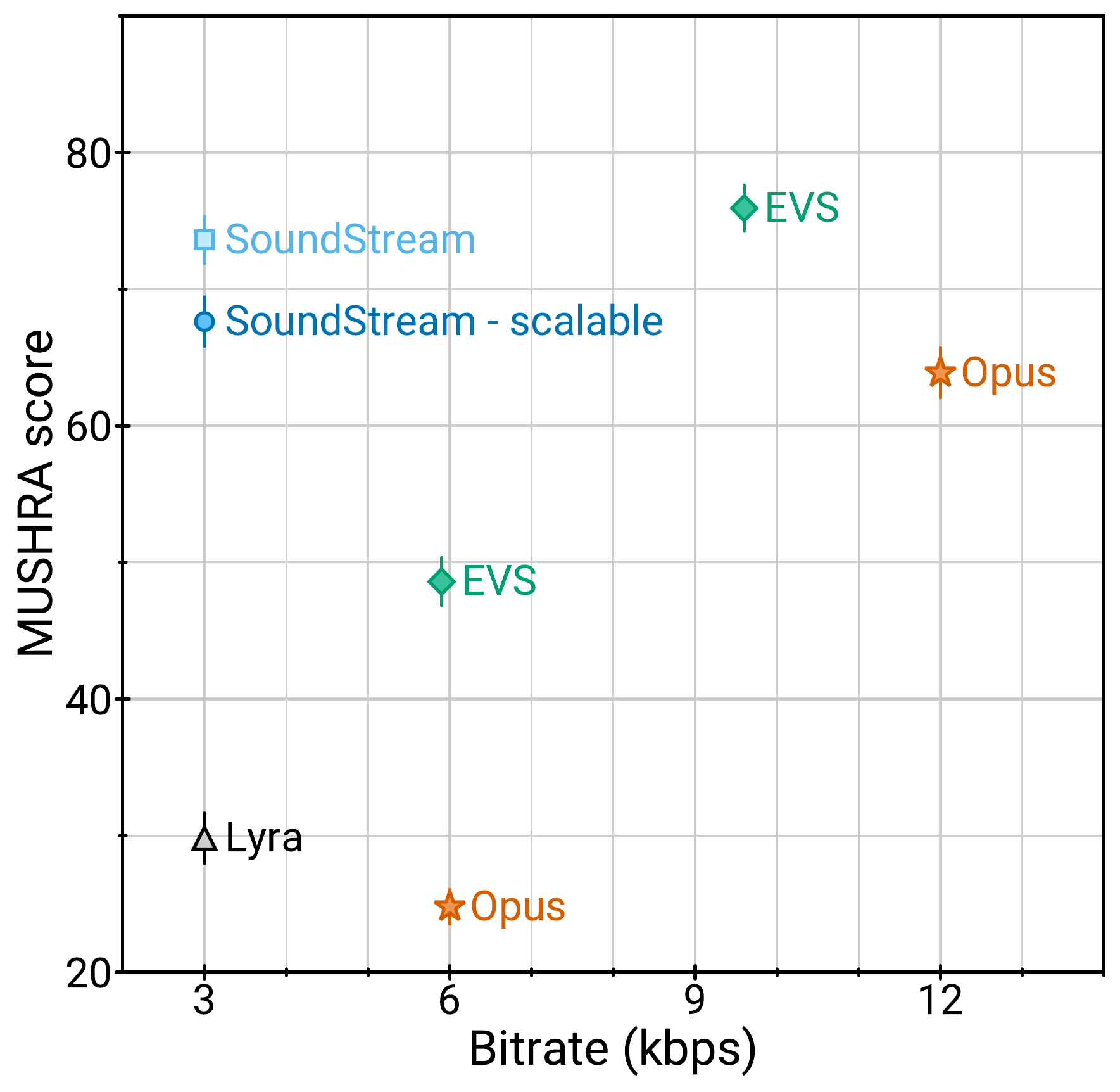}
    \caption{\SoundStream@$3\,$kbps vs. state-of-the-art codecs.}\label{fig:mushra_eval_intro}
\end{figure}

Other solutions adopt ML-based models as an integral part of the audio codec architecture. In these areas, recent advances in text-to-speech (TTS) technology proved to be a key ingredient. For example, WaveNet~\cite{oord2016wavenet}, a strong generative model originally applied to generate speech from text, was adopted as a decoder in a neural codec~\cite{kleijn2018wavenet,garbacea2019vqvae}. Other neural audio codecs adopt different model architectures, e.g., WaveRNN in LPCNet~\cite{valin2019lpcnet} and WaveGRU in Lyra~\cite{kleijn2021lyra}, all targeting speech at low bitrates.

\begin{figure*}
    \centering
    \includegraphics[width=\textwidth]{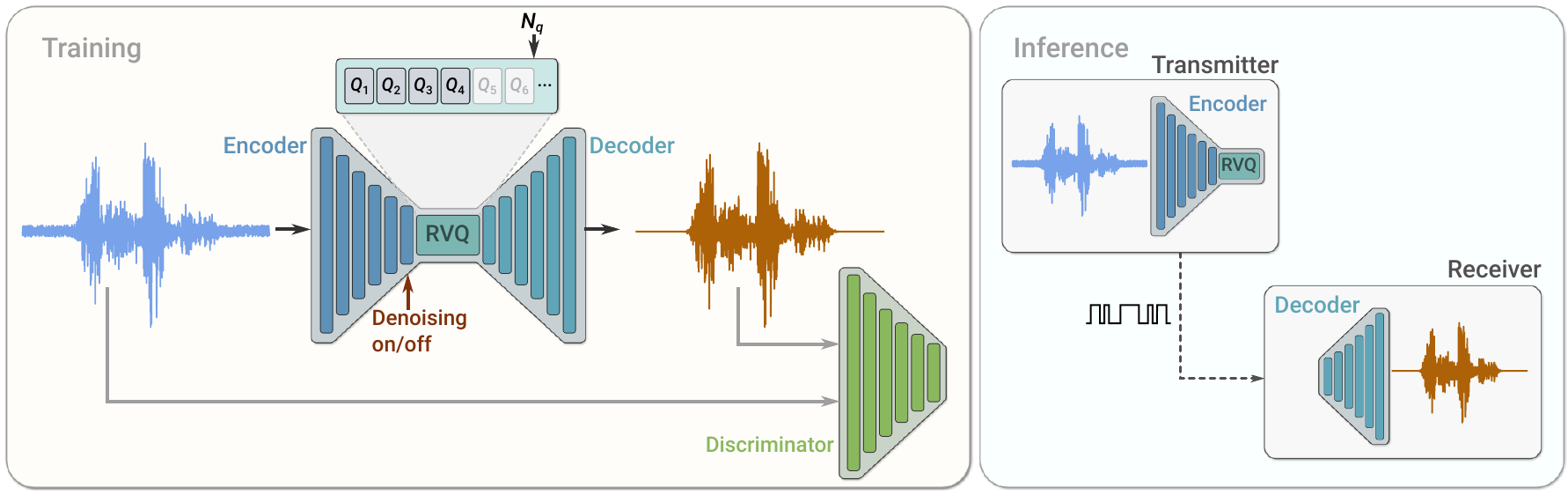}
    \caption{\SoundStream model architecture. A convolutional encoder produces a latent representation of the input audio samples, which is quantized using a variable number $\numselectedquantizers$ of residual vector quantizers (RVQ). During training, the model parameters are optimized using a combination of reconstruction and adversarial losses. An optional conditioning input can be used to indicate whether background noise has to be removed from the audio. When deploying the model, the encoder and quantizer on a transmitter client send the compressed bitstream to a receiver client that can then decode the audio signal.}
    \label{fig:model_architecture}
\end{figure*}

In this paper we propose \SoundStream, a novel audio codec that can compress speech, music and general audio more efficiently than previous codecs, as illustrated in Figure~\ref{fig:mushra_eval_intro}. \SoundStream leverages state-of-the-art solutions in the field of neural audio synthesis, and introduces a new learnable quantization module, to deliver audio at high perceptual quality, while operating at low-to-medium bitrates. Figure~\ref{fig:model_architecture} illustrates the high level model architecture of the codec. A fully convolutional encoder receives as input a time-domain waveform and produces a sequence of embeddings at a lower sampling rate, which are quantized by a residual vector quantizer. A fully convolutional decoder receives the quantized embeddings and reconstructs an approximation of the original waveform. The model is trained end-to-end using both reconstruction and adversarial losses. To this end, one (or more) discriminators are trained jointly, with the goal of distinguishing the decoded audio from the original audio and, as a by-product, provide a space where a feature-based reconstruction loss can be computed. Both the encoder and the decoder only use causal convolutions, so  the overall architectural latency of the model is determined solely by the temporal resampling ratio between the original time-domain waveform and the embeddings. 

In summary, this paper makes the following key contributions:
\begin{itemize}
    \item We propose \SoundStream, a neural audio codec in which all the constituent components (encoder, decoder and quantizer) are trained end-to-end with a mix of reconstruction and adversarial losses to achieve superior audio quality.
    \item We introduce a new residual vector quantizer, and investigate the rate-distortion-complexity trade-offs implied by its design. In addition, we propose a novel ``quantizer dropout'' technique for training the residual vector quantizer, which enables a single model to handle different bitrates.
    \item We demonstrate that learning the encoder brings a very significant coding efficiency improvement, with respect to a solution that adopts mel-spectrogram features.
    \item We demonstrate by means of subjective quality metrics that \SoundStream outperforms both Opus and EVS over a wide range of bitrates.
    \item We design our model to support streamable inference, which can operate at low-latency. When deployed on a smartphone, it runs in real-time on a single CPU thread.
    \item We propose a variant of the \SoundStream codec that performs jointly audio compression and enhancement, without introducing additional latency.
\end{itemize}

\section{Related work}

\textbf{Traditional audio codecs} -- Opus~\cite{opus2012} and EVS~\cite{evs2015} are state-of-the-art audio codecs, which combine traditional coding tools, such as LPC, CELP and MDCT, to deliver high coding efficiency over different content types, bitrates and sampling rates, while ensuring low-latency for real-time audio communications. We compare SoundStream with both Opus and EVS in our subjective evaluation.

\textbf{Audio generative models}  -- Several generative models have been developed for converting text or coded features into audio waveforms. WaveNet~\cite{oord2016wavenet} allows for global and local signal conditioning to synthesize both speech and music. SampleRNN~\cite{mehri2017samplernn} uses recurrent networks in a similar fashion, but it relies on previous samples at different scales. These auto-regressive models deliver very high-quality audio, at the cost of increased computational complexity, since samples are generated one by one. To overcome this issue, Parallel~WaveNet~\cite{oord2017parallelwavenet} allows for parallel computation, yielding considerable speedup during inference. 
Other approaches involve lightweight and sparse models~\cite{kalchbrenner2018wavernn} and networks mimicking the fast Fourier transform as part of the model~\cite{jin2018,valin2019lpcnet}. More recently, generative adversarial models have emerged as a solution able to deliver high-quality audio with a lower computational complexity. MelGAN~\cite{kumar2019melgan} is trained to produce audio waveforms when conditioned on mel-spectrograms, training a multi-scale waveform discriminator together with the generator. HiFiGAN~\cite{kong2020hifigan} takes a similar approach but it applies discriminators to both multiple scales and multiple periods of the audio samples. The design of the decoder and the losses in \SoundStream is based on this class of audio generative models.

\textbf{Audio enhancement} -- Deep neural networks have been applied to different audio enhancement tasks, ranging from denoising~\cite{feng2014dae,pascual2017segan,germain2018senet,rethage2018wavenet,donahue2018denoisinggan} to dereverberation~\cite{ishii2013dae,williamson2017}, lossy coding denoising~\cite{biswas2020aac} and frequency bandwidth extension~\cite{lim2018,li2021seanet}. In this paper we show that it is possible to jointly perform audio enhancement and compression with a single model, without introducing additional latency. 

\textbf{Vector quantization} -- Learning the optimal quantizer is a key element to achieve high coding efficiency. Optimal scalar quantization based on Lloyd's algorithm~\cite{lloyd1982vq} can be extended to a high-dimensional space via the generalized Lloyd algorithm (GLA)~\cite{linde1980lbg}, which is very similar to k-means clustering~\cite{macqueen1967kmeans}. In vector quantization~\cite{gray1984vq}, a point in a high-dimensional space is mapped onto a discrete set of code vectors. Vector quantization has been commonly used as a building block of traditional audio codecs~\cite{makhoul1985vector}. For example, CELP~\cite{schroder1985celp} adopts an excitation signal encoded via a vector quantizer codebook. More recently, vector quantization has been applied in the context of neural network models to compress the latent representation of input features. For example, in variational autoencoders, vector quantization has been used to generate images~\cite{vqvae,vqvae2} and music~\cite{dieleman2018music, dhariwal2020jukebox}.
Vector quantization can become prohibitively expensive, as the size of the codebook grows exponentially when rate is increased. For this reason, structured vector quantizers~\cite{juang1982multiple,vasuki2006vq} (e.g., residual, product, lattice vector quantizers, etc.) have been proposed to obtain a trade-off between computational complexity and coding efficiency in traditional codecs. In \SoundStream, we extend the learnable vector quantizer of VQ-VAE \cite{vqvae} and introduce a residual (a.k.a. multi-stage) vector quantizer, which is learned end-to-end with the rest of the model. To the best of the authors knowledge, this is the first time that this form of vector quantization is used in the context of neural networks and trained end-to-end with the rest of the model.

\textbf{Neural audio codecs} -- End-to-end neural audio codecs rely on data-driven methods to learn efficient audio representations, instead of relying on handcrafted signal processing components. Autoencoder networks with quantization of hidden features were applied to speech coding early on~\cite{morishima1990speechcoding}. More recently, a more sophisticated deep convolutional network for speech compression was described in~\cite{kankanahalli2018speechdnn}. Efficient compression of audio using neural networks has been demonstrated in several works, mostly targeting speech coding at low bitrates. %
A VQ-VAE speech codec was proposed in~\cite{garbacea2019vqvae}, operating at $1.6\,$kbps. Lyra~\cite{kleijn2021lyra} is a generative model that encodes quantized mel-spectrogram features of speech, which are decoded with an auto-regressive WaveGRU model to achieve state-of-the-art results at $3\,$kbps. A very low-bitrate codec was proposed in~\cite{polyak2021speech} by decoding speech representations obtained via self-supervised learning.
An end-to-end audio codec targeting general audio at high bitrates (i.e., above $64\,$kbps) was proposed in~\cite{zhen2019cascaded}. The model architecture adopts a residual coding pipeline, which consists of multiple autoencoding modules and a psycho-acoustic model is used to drive the loss function during training.

Unlike~\cite{polyak2021speech} which specifically targets speech by combining speaker, phonetic and pitch embeddings, \SoundStream does not make assumptions on the nature of the signal it encodes, and thus works for diverse audio content types. While~\cite{kleijn2021lyra} learns a decoder on fixed features, \SoundStream is trained in an end-to-end fashion. Our experiments (see Section \ref{sec:experiments}) show that learning the encoder increases the audio quality substantially. \SoundStream achieves bitrate scalability, i.e., the ability of a single model to operate at different bitrates at no additional cost, thanks to its residual vector quantizer and to our original quantizer dropout training scheme (see Section \ref{subsec:vector_quantizer}). This is unlike \cite{kankanahalli2018speechdnn} and \cite{zhen2019cascaded} which enforce a specific bitrate during training and require training a different model for each target bitrate. A single \SoundStream model is able to compress speech, music and general audio, while operating at a $24\,$kHz sampling rate and low-to-medium bitrates ($3\,$kbps to $18\,$kbps in our experiments), in real time on a smartphone CPU. This is the first time that a neural audio codec is shown to outperform state-of-the-art codecs like Opus and EVS over this broad range of bitrates.

\textbf{Joint compression and enhancement} --  Recent work has explored joint compression and enhancement. The work in~\cite{casebeer2021} trains a speech enhancement system with a quantized bottleneck. Instead, \SoundStream integrates a time-dependent conditioning layer, which allows for real-time controllable denoising. As we design \SoundStream as a general-purpose audio codec, controlling when to denoise allows for encoding acoustic scenes and natural sounds that would be otherwise removed.

\section{Model}
We consider a single channel recording $\waveform \in \mathbb{R}^{\numsamples}$, sampled at $\samplerate$. The \SoundStream model consists of a sequence of three building blocks, as illustrated in Figure~\ref{fig:model_architecture}: 
\begin{itemize}
    \item an encoder, which maps $\waveform$ to a sequence of embeddings (see Section~\ref{subsec:encoder}),
    \item a residual vector quantizer, which replaces each embedding by the sum of vectors from a set of finite codebooks, thus compressing the representation with a target number of bits (see Section~\ref{subsec:vector_quantizer}),
    \item a decoder, which produces a lossy reconstruction $\waveformhat \in \mathbb{R}^{\numsamples}$ from quantized embeddings (see Section~\ref{subsec:decoder}). 
\end{itemize}
The model is trained end-to-end together with a discriminator (see Section~\ref{subsec:discriminator}), using the mix of adversarial and reconstruction losses described in Section~\ref{subsec:losses}. Optionally, a conditioning signal can be added, which determines whether denoising is applied at the encoder or decoder side, as detailed in Section~\ref{subsec:conditioning}.

\begin{figure*}
    \centering
    \includegraphics[width=\textwidth]{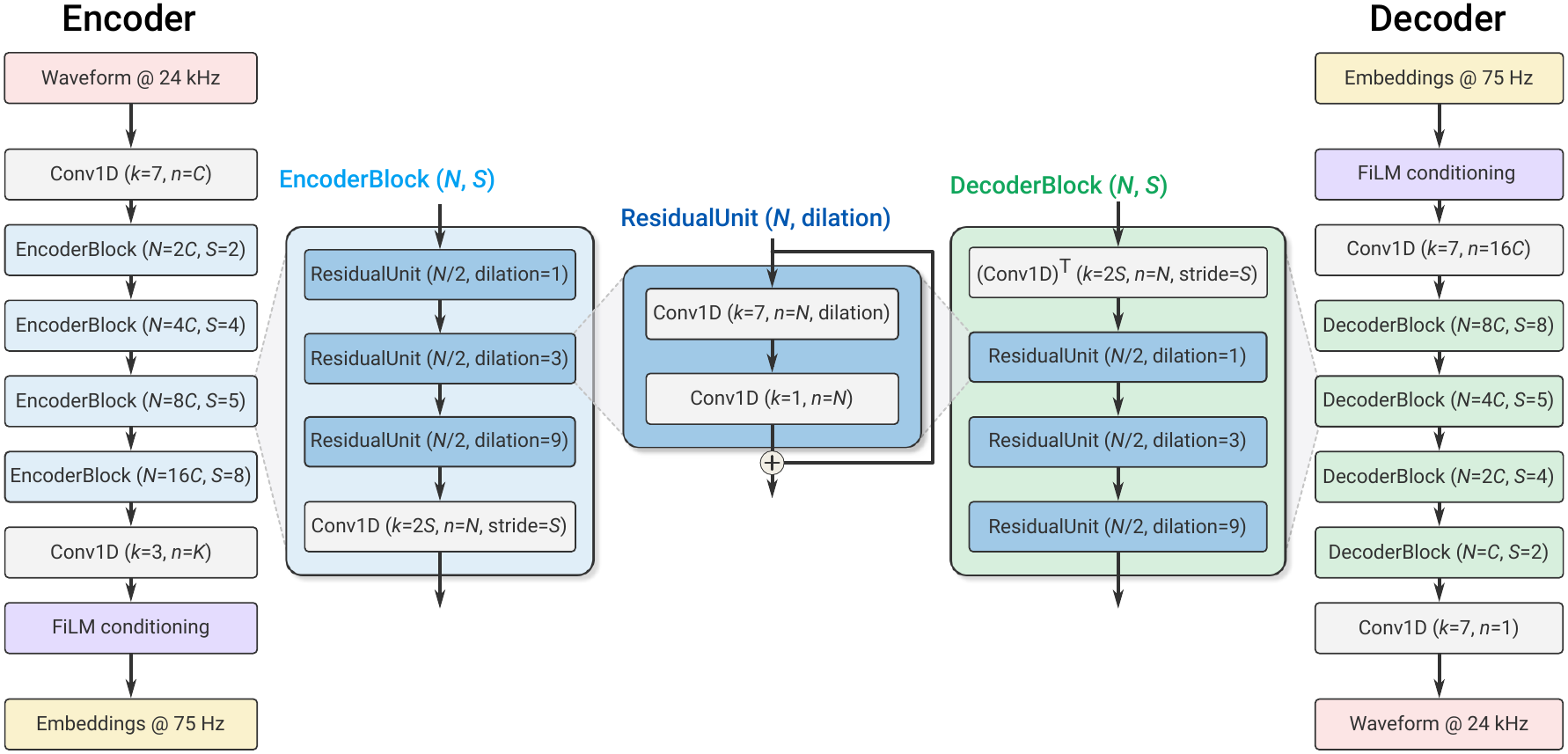}
    \caption{Encoder and decoder model architecture.}
    \label{fig:encoder_decoder_architecture}
\end{figure*}

\subsection{Encoder architecture}\label{subsec:encoder}
The encoder architecture is illustrated in Figure~\ref{fig:encoder_decoder_architecture} and follows the same structure as the \textit{streaming SEANet} encoder described in~\cite{li2021seanet}, but without skip connections. It consists of a 1D convolution layer (with $\baseconvdepth_{\enc}$ channels), followed by $\numblocks_{\enc}$ convolution blocks. Each of the blocks consists of three residual units, containing dilated convolutions with dilation rates of 1, 3, and 9, respectively, followed by a down-sampling layer in the form of a strided convolution. The number of channels is doubled whenever down-sampling, starting from $\baseconvdepth_{\enc}$. A final 1D convolution layer with a kernel of length 3 and a stride of 1 is used to set the dimensionality of the embeddings to $\dims$. To guarantee real-time inference, all convolutions are \emph{causal}. This means that padding is only applied to the past but not the future in both training and offline inference, whereas no padding is used in streaming inference. We use the ELU activation \cite{elu} and we do not apply any normalization. The number $\numblocks_{\enc}$ of convolution blocks and the corresponding striding sequence determines the temporal resampling ratio between the input waveform and the embeddings. For example, when $\numblocks_{\enc} = 4$ and using $(2, 4, 5, 8)$ as strides, one embedding is computed every $\ratio=2\cdot 4 \cdot 5 \cdot 8 = 320$ input samples. Thus, the encoder outputs $\enc(x) \in \mathbb{R}^{\numframes \times \dims}$, with $\numframes =  \numsamples / \ratio$.

\subsection{Decoder architecture}\label{subsec:decoder}
The decoder architecture follows a similar design, as illustrated in Figure~\ref{fig:encoder_decoder_architecture}. A 1D convolution layer is followed by a sequence of $\numblocks_{\dec}$ convolution blocks. The decoder block mirrors the encoder block, and consists of a transposed convolution for up-sampling followed by the same three residual units. We use the same strides as the encoder, but in reverse order, to reconstruct a waveform with the same resolution as the input waveform. The number of channels is halved whenever up-sampling, so that the last decoder block outputs $\baseconvdepth_{\dec}$ channels. A final 1D convolution layer with one filter, a kernel of size 7 and stride 1 projects the embeddings back to the waveform domain to produce $\waveformhat$. In Figure~\ref{fig:encoder_decoder_architecture}, the same number of channels in both the encoder and the decoder is controlled by the same parameter, i.e., $\baseconvdepth_{\enc} = \baseconvdepth_{\dec} = \baseconvdepth$. We also investigate cases in which $\baseconvdepth_{\enc} \ne \baseconvdepth_{\dec}$, which results in a computationally lighter encoder and a heavier decoder, or vice-versa (see Section \ref{subsec:ablations}).

\subsection{Residual Vector Quantizer:}\label{subsec:vector_quantizer}

\begin{algorithm}[t]
\DontPrintSemicolon
\caption{Residual Vector Quantization}\label{algo:rvq}
\SetKwInput{Input}{Input}
\SetKwInput{Output}{Output}
\Input{$y = \enc(x)$ the output of the encoder, vector quantizers $Q_i$ for $i=1..\numquantizers$}
\Output{the quantized $\hat{y}$} %
$\hat{y} \gets 0.0$\;
${\rm residual} \gets y$\;

\For{$i=1$ to $N_q$}{
$\hat{y} \pluseq Q_i({\rm residual})$\;
${\rm residual} \minuseq Q_i({\rm residual})$\;
}
\textbf{return} $\hat{y}$
\end{algorithm}

The goal of the quantizer is to compress the output of the encoder $\enc(x)$ to a target bitrate $\bitrate$, expressed in bits/second~(bps). In order to train \SoundStream in an end-to-end fashion, the quantizer needs to be jointly trained with the encoder and the decoder by backpropagation. The vector quantizer (VQ) proposed in~\cite{vqvae, vqvae2} in the context of VQ-VAEs meets this requirement. This vector quantizer learns a codebook of $\codebooksize$ vectors to encode each $\dims$-dimensional frame of $\enc(x)$. The encoded audio $\enc(x) \in \mathbb{R}^{\numframes \times \dims}$ is then mapped to a sequence of one-hot vectors of shape $\numframes \times \codebooksize$, which can be represented using $\numframes\log_{2}\codebooksize$ bits. 

\noindent\textbf{Limitations of Vector Quantization} -- As a concrete example, let us consider a codec targeting a bitrate $\bitrate = 6000\,$bps. When using a striding factor $\ratio = 320$, each second of audio at sampling rate $\samplerate = 24000\,$Hz is represented by $\numframes = 75$~frames at the output of the encoder. This corresponds to $\rateperembedding = 6000/75 = 80$ bits allocated to each frame. 
Using a plain vector quantizer, this requires storing a codebook with $\codebooksize = 2^{80}$ vectors, which is obviously unfeasible.

\noindent\textbf{Residual Vector Quantizer} -- To address this issue we adopt a Residual Vector Quantizer (a.k.a. multi-stage vector quantizer~\cite{vasuki2006vq}), which cascades $\numquantizers$ layers of VQ as follows. The unquantized input vector is passed through a first VQ and quantization residuals are computed. The residuals are then iteratively quantized by a sequence of additional $\numquantizers - 1$ vector quantizers, as described in~Algorithm \ref{algo:rvq}. The total rate budget $\rm$ is uniformly allocated to each VQ, i.e., $\rateperembedding_i = \rateperembedding / \numquantizers = \log_2 \codebooksize$. For example, when using $\numquantizers = 8$, each quantizer uses a codebook of size $\codebooksize = 2^{\rateperembedding / \numquantizers} = 2^{80 / 8} = 1024$. For a target rate budget $\rateperembedding$, the parameter $\numquantizers$ controls the tradeoff between computational complexity and coding efficiency, which we investigate in Section~\ref{subsec:ablations}. 

The codebook of each quantizer is trained with exponential moving average updates, following the method proposed in VQ-VAE-2~\cite{vqvae2}. To improve the usage of the codebooks we use two additional methods. First, instead of using a random initialization for the codebook vectors, we run the k-means algorithm on the first training batch and use the learned centroids as initialization. This allows the codebook to be close to the distribution of its inputs and improves its usage. Second, as proposed in~\cite{dhariwal2020jukebox}, when a codebook vector has not been assigned any input frame for several batches, we replace it with an input frame randomly sampled within the current batch. More precisely, we track the exponential moving average of the assignments to each vector (with a decay factor of $0.99$) and replace the vectors of which this statistic falls below $2$. 

\noindent\textbf{Enabling bitrate scalability with quantizer dropout} -- 
Residual vector quantization provides a convenient framework for controlling the bitrate. For a fixed size $\codebooksize$ of each codebook, the number of VQ layers $\numquantizers$ determines the bitrate. Since the vector quantizers are trained jointly with the encoder/decoder, in principle a different \SoundStream model should be trained for each target bitrate. Instead, having a single \textit{bitrate scalable} model that can operate at several target bitrates is much more practical, since this reduces the memory footprint needed to store model parameters both at the encoder and decoder side.

To train such a model, we modify Algorithm~\ref{algo:rvq} in the following way: for each input example, we sample $\numselectedquantizers$ uniformly at random in $[1;\numquantizers]$ and only use quantizers $Q_i$ for $i=1\dots\numselectedquantizers$. This can be seen as a form of structured dropout~\cite{srivastava2014dropout} applied to quantization layers. Consequently, the model is trained to encode and decode audio for all target bitrates corresponding to the range $\numselectedquantizers=1\dots\numquantizers$. During inference, the value of $\numselectedquantizers$ is selected based on the desired bitrate.
 Previous models for neural compression have relied on product quantization (wav2vec 2.0 \cite{wav2vec2}), or on concatenating the output of several VQ layers \cite{kleijn2018wavenet, garbacea2019vqvae}. With such approaches, changing the bitrate requires either changing the architecture of the encoder and/or the decoder, as the dimensionality changes, or retraining an appropriate codebook. A key advantage of our residual vector quantizer is that the dimensionality of the embeddings does not change with the bitrate. Indeed, the additive composition of the outputs of each VQ layer progressively refines the quantized embeddings, while keeping the same shape. Hence, no architectural changes are needed in neither the encoder nor the decoder to accommodate different bitrates. 
In Section~\ref{subsec:bitrate_scalability}, we show that this method allows one to train a single \SoundStream model, which matches the performance of models trained specifically for a given bitrate.

\subsection{Discriminator architecture}\label{subsec:discriminator}
To compute the adversarial losses described in Section~\ref{subsec:losses}, we define two different discriminators: i) a wave-based discriminator, which receives as input a single waveform; ii) an STFT-based discriminator, which receives as input the complex-valued STFT of the input waveform, expressed in terms of real and imaginary parts. Since both discriminators are fully convolutional, the number of logits in the output is proportional to the length of the input audio. 

For the wave-based discriminator, we use the same multi-resolution convolutional discriminator proposed in~\cite{kumar2019melgan} and adopted in~\cite{tagliasacchi2020seanet}. Three structurally identical models are applied to the input audio at different resolutions: original, 2-times down-sampled, and 4-times down-sampled. Each single-scale discriminator consists of an initial plain convolution followed by four grouped convolutions, each of which has a group size of 4, a down-sampling factor of 4, and a channel multiplier of 4 up to a maximum of 1024 output channels. They are followed by two more plain convolution layers to produce the final output, i.e., the logits. 

\begin{figure}
    \centering
    \includegraphics[width=\columnwidth]{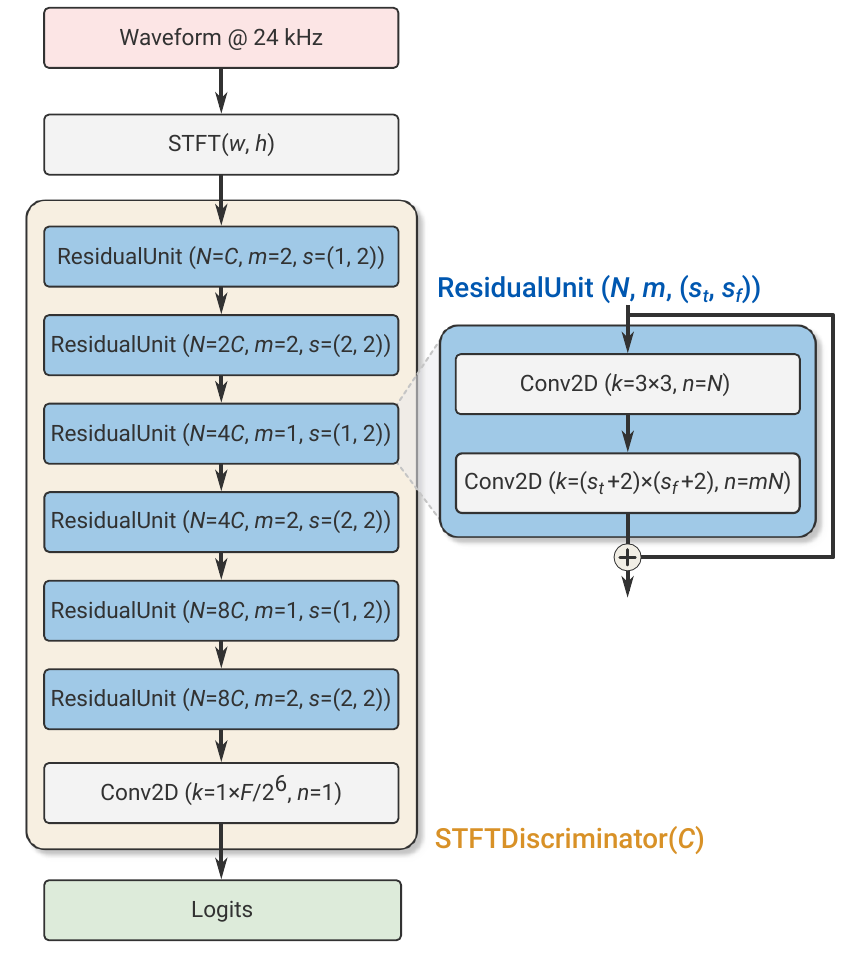}
    \caption{STFT-based discriminator architecture.}
    \label{fig:stft_discriminator_architecture}
\end{figure}

The STFT-based discriminator is illustrated in Figure~\ref{fig:stft_discriminator_architecture} and operates on a single scale, computing the STFT with a window length of $\windowlen=1024$ samples and a hop length of $\hoplen=256$~samples. A 2D-convolution (with kernel size $7\times7$ and 32 channels) is followed by a sequence of residual blocks. Each block starts with a $3\times3$ convolution, followed by a $3\times4$ or a $4\times4$ convolution, with strides equal to $(1, 2)$ or $(2, 2)$, where $(s_t, s_f)$ indicates the down-sampling factor along the time axis and the frequency axis. We alternate between $(1, 2)$ and $(2, 2)$ strides, for a total of 6 residual blocks. The number of channels is progressively increased with the depth of the network. At the output of the last residual block, the activations have shape $\numsamples / (\hoplen \cdot 2^3) \times \numbins / 2^6$, where $\numsamples$ is the number of samples in the time domain and $\numbins = \windowlen / 2$ is the number of frequency bins. The last layer aggregates the logits across the (down-sampled) frequency bins with a fully connected layer (implemented as a $1\times \numbins / 2^6 $ convolution), to obtain a 1-dimensional signal in the (down-sampled) time domain.

\subsection{Training objective}\label{subsec:losses}
Let $\G(x) = \dec (\quant (\enc(\waveform))$ denote the \SoundStream generator, which processes the input waveform $x$ through the encoder, the quantizer and the decoder, and $\waveformhat = \G(\waveform)$ be the decoded waveform. We train~\SoundStream with a mix of losses to achieve both signal reconstruction fidelity  and perceptual quality, following the principles of the perception-distortion trade-off discussed in~\cite{blau2018perception}. 

The adversarial loss is used to promote perceptual quality and it is defined as a hinge loss over the logits of the discriminator, averaged over multiple discriminators and over time.
More formally, let $k\in\{0,\dots,\numdiscr\}$ index over the individual discriminators, where $k=0$ denotes the STFT-based discriminator and $k\in \{1,\dots,\numdiscr\}$ the different resolutions of the waveform-based discriminator ($\numdiscr=3$ in our case). Let $\numdiscrout$ denote the number of logits at the output of the $k$-th discriminator along the time dimension. The discriminator is trained to classify original vs. decoded audio, by minimizing
\begin{align}
    \nonumber
    \Loss_\D =~& \Exp_{\waveform} \left [\frac{1}{\numdiscr}\sum_{k} \frac{1}{\numdiscrout}\sum_{t} \max\Big(0, 1 - \D_{k,t}(\waveform)\Big) \right ] + \\  %
    & \Exp_{\waveform} \left [ \frac{1}{\numdiscr}\sum_{k} \frac{1}{\numdiscrout}\sum_{t} \max\Big(0, 1 + \D_{k,t}(\G(\waveform))\Big) \right ],  %
\end{align}
while the adversarial loss for the generator is
\begin{equation}
    \Loss_\G^{\text{adv}} = \Exp_{x} \left [ \frac{1}{\numdiscr}\sum_{k,t} \frac{1}{\numdiscrout} \max\Big(0, 1 - \D_{k,t}(\G(x))\Big) \right ].  %
\end{equation}

To promote fidelity of the decoded signal $\waveformhat$ with respect to the original $x$ we adopt two additional losses: i) a ``feature'' loss $\Loss_\G^{\text{feat}}$, computed in the feature space defined by the discriminator(s)~\cite{kumar2019melgan}; ii) a multi-scale spectral reconstruction loss $\Loss_\G^{\text{rec}}$~\cite{engel2020ddsp}. 

More specifically, the feature loss is computed by taking the average absolute difference between the discriminator's internal layer outputs for the generated audio and those for the corresponding target audio.
\begin{equation}
    \Loss_\G^{\text{feat}} = \Exp_{x} \left [ \frac{1}{\numdiscr \numlayers}\sum_{k,l} \frac{1}{T_{k,l}} \sum_{t} \left| \D_{k,t}^{(l)}(\waveform) - \D_{k,t}^{(l)}(\G(\waveform)) \right| \right ],
\end{equation}
where $L$ is the number of internal layers, $\D_{k,t}^{(l)}$ ($l\in\{1,\dots,\numlayers\}$) is the $t$-th output of layer $l$ of discriminator $k$, and $T_{k,l}$ denotes the length of the layer in the time dimension.

The multi-scale spectral reconstruction loss follows the specifications described in~\cite{gritsenko2020spectral}:
\begin{align}
    \Loss_\G^{\text{rec}} = \sum_{s \in {2^{6}, \dots, 2^{11}}}& \sum_t \|\SPEC^s_t(x) - \SPEC^s_t(\G(\waveform)) \|_1 + \\ &\alpha_s \sum_t \|\log \SPEC^s_t(\waveform) - \log \SPEC^s_t(\G(\waveform)) \|_2,
\end{align}
where $\SPEC^s_t(x)$ denotes the $t$-th frame of a 64-bin mel-spectrogram computed with window length equal to $s$ and hop length equal to $s/4$. We set $\alpha_s = \sqrt{s / 2}$ as in~\cite{gritsenko2020spectral}.

The overall generator loss is a weighted sum of the different loss components:
\begin{equation}
    \Loss_G = \lambda_{\text{adv}}\Loss_G^{\text{adv}} + \lambda_{\text{feat}} \cdot \Loss_G^{\text{feat}} + \lambda_{\text{rec}} \cdot \Loss_G^{\text{rec}}.
\end{equation}
In all our experiments we set $\lambda_{\text{adv}} = 1$, $\lambda_{\text{feat}} = 100$ and $\lambda_{\text{rec}} = 1$.

\subsection{Joint compression and enhancement}\label{subsec:conditioning}

In traditional audio processing pipelines, compression and enhancement are typically performed by different modules. For example, it is possible to apply an audio enhancement algorithm at the transmitter side, before audio is compressed, or at the receiver side, after audio is decoded. In this setup, each processing step contributes to the end-to-end latency, e.g., due to buffering the input audio to the expected frame length determined by the specific algorithm adopted. Conversely, we design \SoundStream in such a way that compression and enhancement can be carried out jointly by the same model, without increasing the overall latency. 

The nature of the enhancement can be determined by the choice of the training data. As a concrete example, in this paper we show that it is possible to combine compression with background noise suppression. More specifically, we train a model in such a way that one can flexibly enable or disable denoising at inference time, by feeding a  conditioning signal that represents the two modes (denoising enabled or disabled). To this end, we prepare the training data to consist of tuples of the form: $(\inputsamples, \targetsamples, \denoise)$. When $\denoise=\false$, $\targetsamples = \inputsamples$; when $\denoise=\true$, $\targetsamples$ contain the clean speech component of the corresponding $\inputsamples$. Hence, the network is trained to reconstruct noisy speech if the conditioning signal is disabled, and to produce a clean version of the noisy input if it is enabled. Note that when $\inputsamples$ consist of clean audio (speech or music), $\targetsamples = \inputsamples$ and $\denoise$ can be either $\true$ or $\false$. This is done to prevent \SoundStream from adversely affecting clean audio when denoising is enabled.

To process the conditioning signal, we use Feature-wise Linear Modulation (FiLM) layers~\cite{perez2018film} in between residual units, which take network features as inputs and transform them as
\begin{equation}
    \activationtrans = \filmmultcoeff \activation + \filmaddcoeff,
\end{equation}
where $\activation$ is the $n^{\rm th}$ activation in the $c^{\rm th}$ channel. The coefficients $\filmmultcoeff$ and $\filmaddcoeff$ are computed by a linear layer that takes as input a (potentially time-varying) two-dimensional one-hot encoding that determines the denoising mode. This allows one to adjust the level of denoising over time. 

In principle, FiLM layers can be used anywhere throughout the encoder and decoder architecture. However, in our preliminary experiments, we found that applying conditioning at the bottleneck either at the encoder or at the decoder side (as illustrated in Figure~\ref{fig:encoder_decoder_architecture}) was effective and no further improvements were observed by applying FiLM layers at different depths. In Section~\ref{subsec:joint_compression_enhancement}, we quantify the impact of enabling denoising at either the encoder or decoder side both in terms of audio quality and bitrate.

\section{Evaluation setup}\label{sec:experiments}
\subsection{Datasets}\label{subsec:datasets}
We train SoundStream on three types of audio content: clean speech, noisy speech and music, all at $24\,$kHz sampling rate. For clean speech, we use the LibriTTS dataset~\cite{zen2019libritts}. For noisy speech, we synthesize samples by mixing speech from LibriTTS with noise from Freesound~\cite{fonseca2017freesound}. We apply peak normalization to randomly selected crops of 3 seconds and adjust the mixing gain of the noise component sampling uniformly in the interval $[-30 \, \text{dB}, 0 \, \text{dB}]$. For music, we use the MagnaTagATune dataset~\cite{law2009magnatagatune}. 
We evaluate our models on disjoint test splits of the datasets above. In addition, we collected a real-world dataset, which contains both near-field and far-field (reverberant) speech, with background noise in some of the examples. 
Unless stated otherwise, objective and subjective metrics are computed on a set of 200 audio clips 2-4 seconds long, with 50 samples from each of the four datasets listed above (i.e., clean speech, noisy speech, music, noisy/reverberant speech).

\subsection{Evaluation metrics}
To evaluate \SoundStream, we perform subjective evaluations by human raters. We have chosen a crowd-sourced methodology inspired by MUSHRA~\cite{BS1534}, with a hidden reference but no lowpass-filtered anchor. Each of the 200 samples of the evaluation dataset, which include clean, noisy and reverberant speech, as well as music, was rated 20 times. The raters were required to be native English speakers and be using headphones. Additionally, to avoid noisy data, a post-screening was put in place to exclude ratings by listeners who rated the reference below 90 more than 20\% of the time or rated non-reference samples above 90 more than 50\% of the time.

For development and hyperparameter selection, we rely on computational, objective metrics. Numerous metrics have been developed in the past for assessing the perceived similarity between a reference and a processed audio signal. The \mbox{ITU-T} standards PESQ~\cite{pesq} and its replacement POLQA~\cite{polqa3} are commonly used metrics. However, both are inconvenient to use owing to licensing restrictions. We choose the freely available and recently open-sourced ViSQOL~\cite{hines2012visqol, chinen2020visqol} metric, which has previously shown comparable performance to POLQA.  In early experiments, we found this metric to be strongly correlated with subjective evaluations. We thus use it for model selection and ablation studies.

\begin{figure*}[t]
    \centering
    \begin{subfigure}[b]{0.32 \textwidth}
        \includegraphics[width=1.0\textwidth]{figures/mushra_dots_3kbps.pdf}
        \caption{Low bitrate.}
        \label{fig:mushra_3kbps}
    \end{subfigure}
    ~ %
    \begin{subfigure}[b]{0.32 \textwidth}
        \includegraphics[width=1.0 \textwidth]{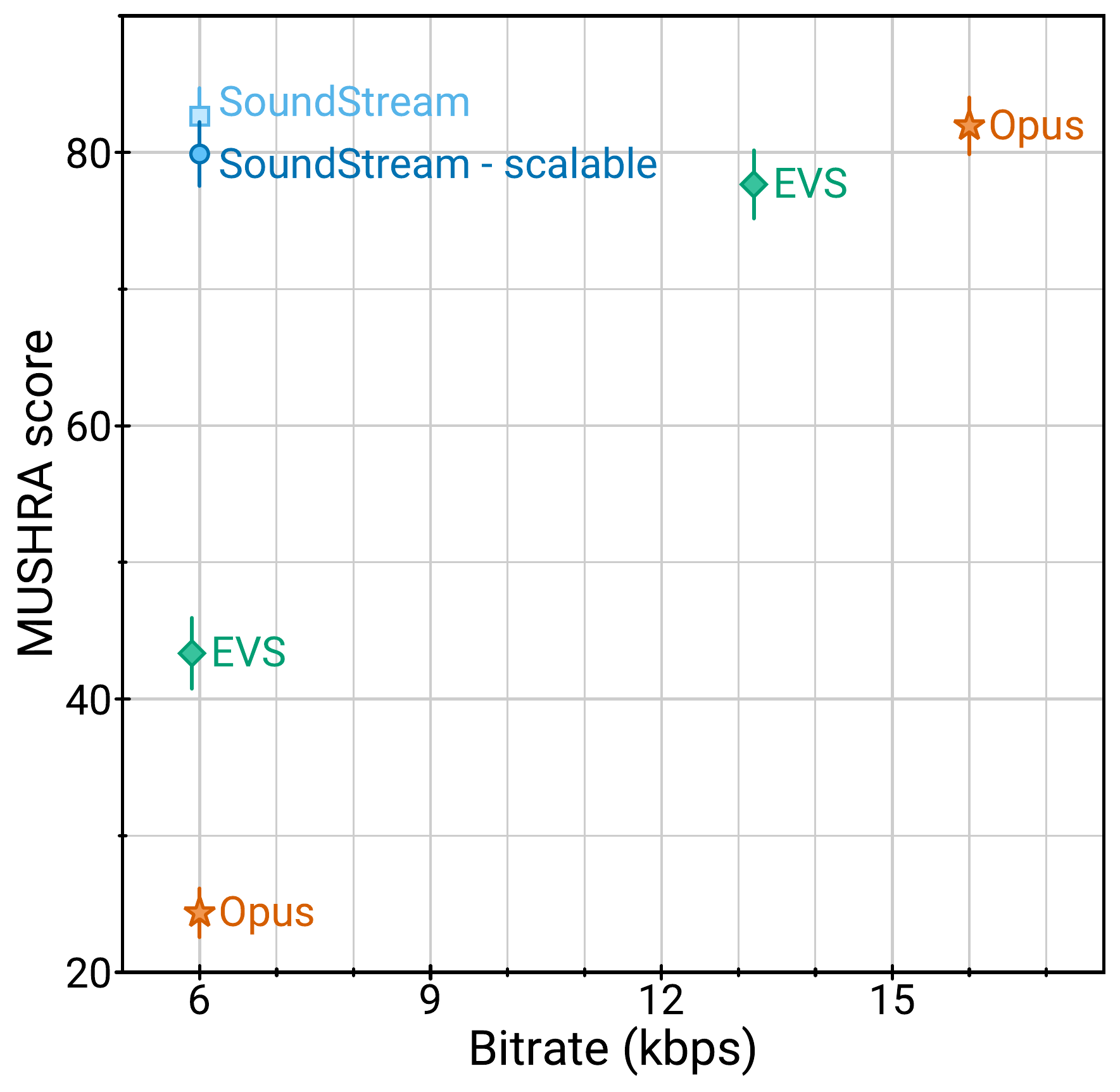}
        \caption{Medium bitrate.}
        \label{fig:mushra_6kbps}
    \end{subfigure}
    \begin{subfigure}[b]{0.32 \textwidth}
        \includegraphics[width=1.0 \textwidth]{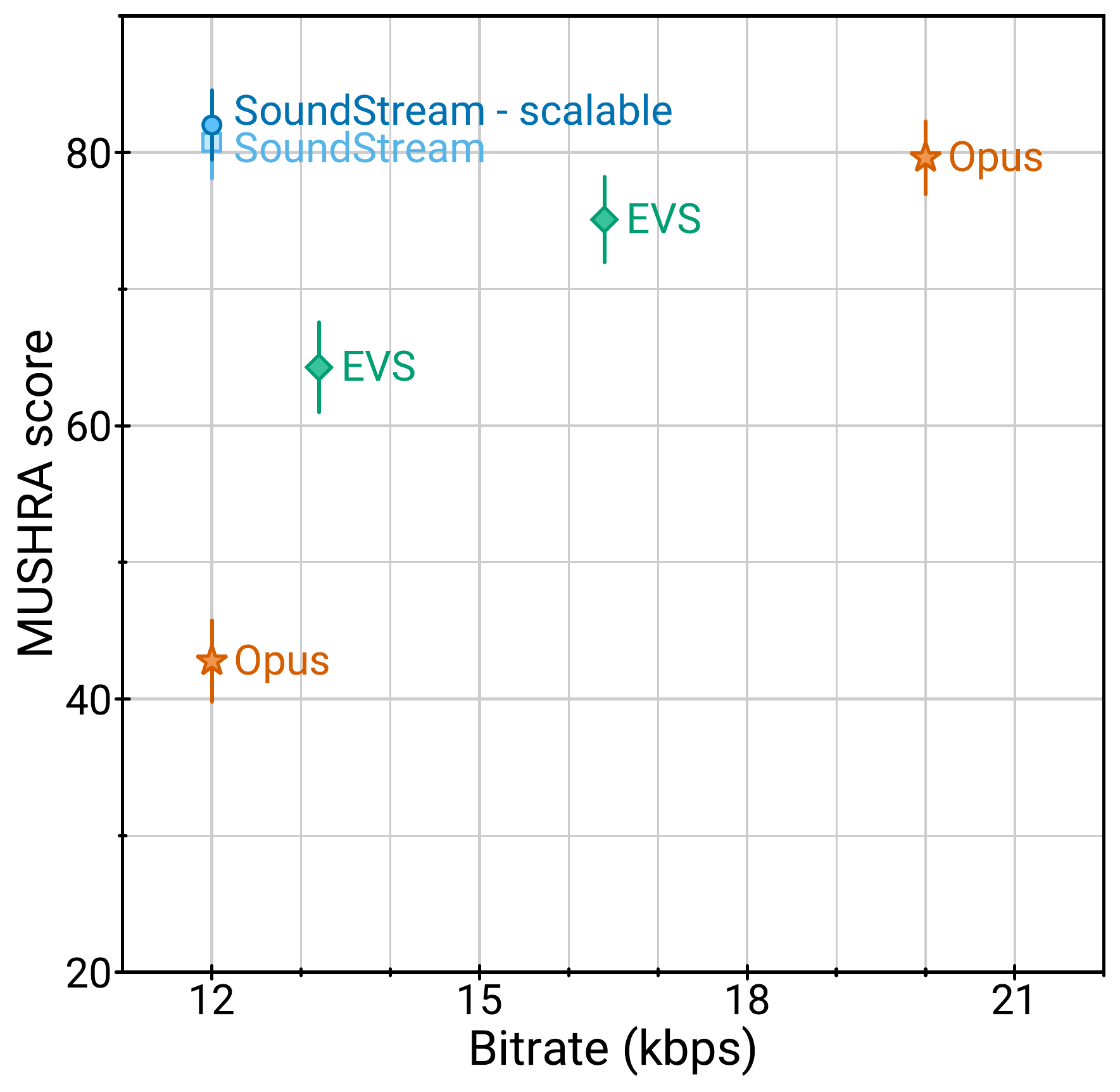}
        \caption{High bitrate.}
        \label{fig:mushra_12kbps}
    \end{subfigure}    
    \caption{Subjective evaluation results. Error bars denote $95\%$ confidence intervals.}\label{fig:mushra_eval}
\end{figure*}

\begin{figure*}[t]
    \centering
    \begin{subfigure}[b]{0.32 \textwidth}
        \includegraphics[width=1.0\columnwidth]{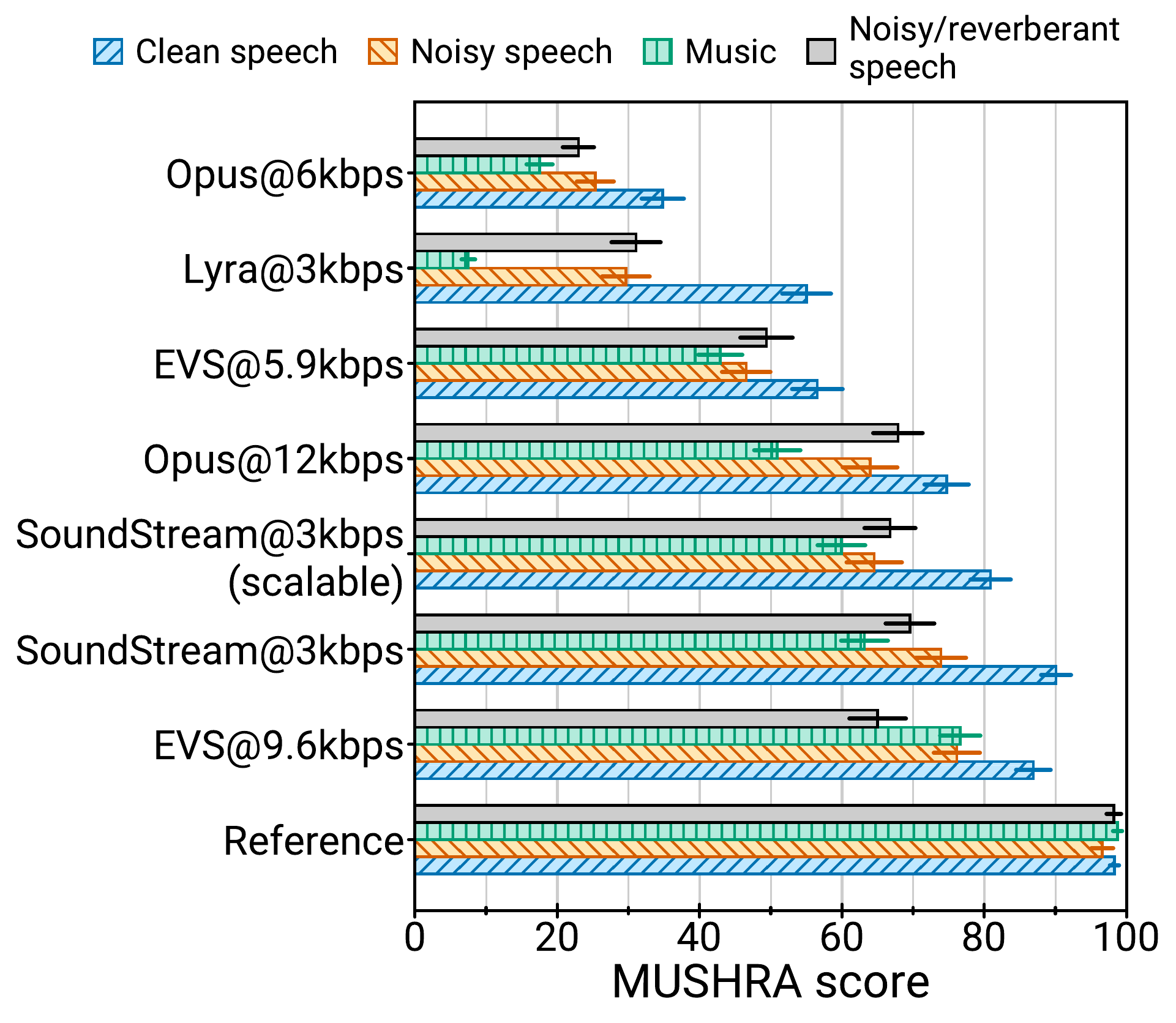}
        \caption{Low bitrate.}
        \label{fig:mushra_3kbps_by_dataset}
    \end{subfigure}
    ~ %
    \begin{subfigure}[b]{0.32 \textwidth}
        \includegraphics[width=1.0 \textwidth]{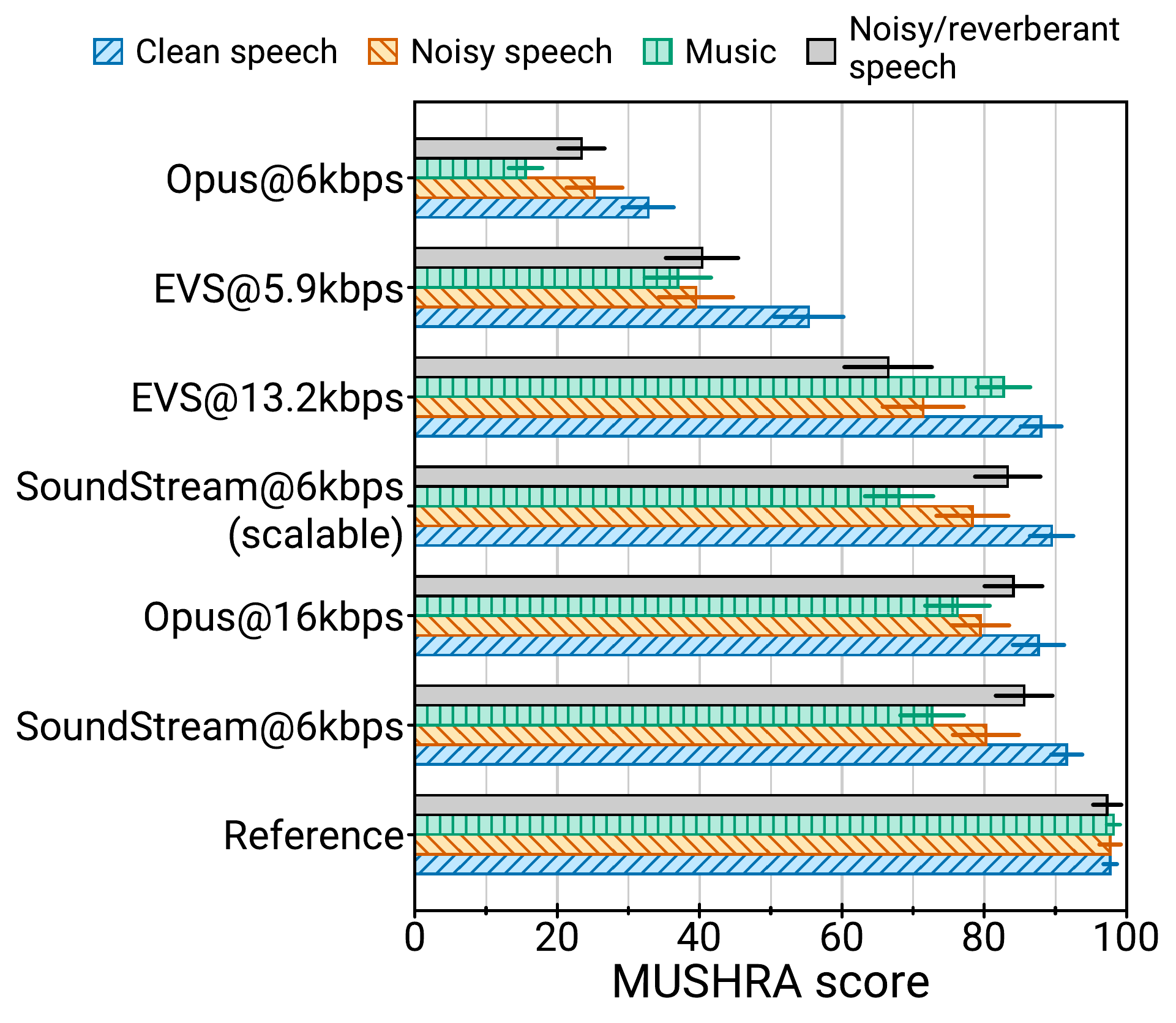}
        \caption{Medium bitrate.}
        \label{fig:mushra_6kbps_by_dataset}
    \end{subfigure}
    \begin{subfigure}[b]{0.32 \textwidth}
        \includegraphics[width=1.0 \textwidth]{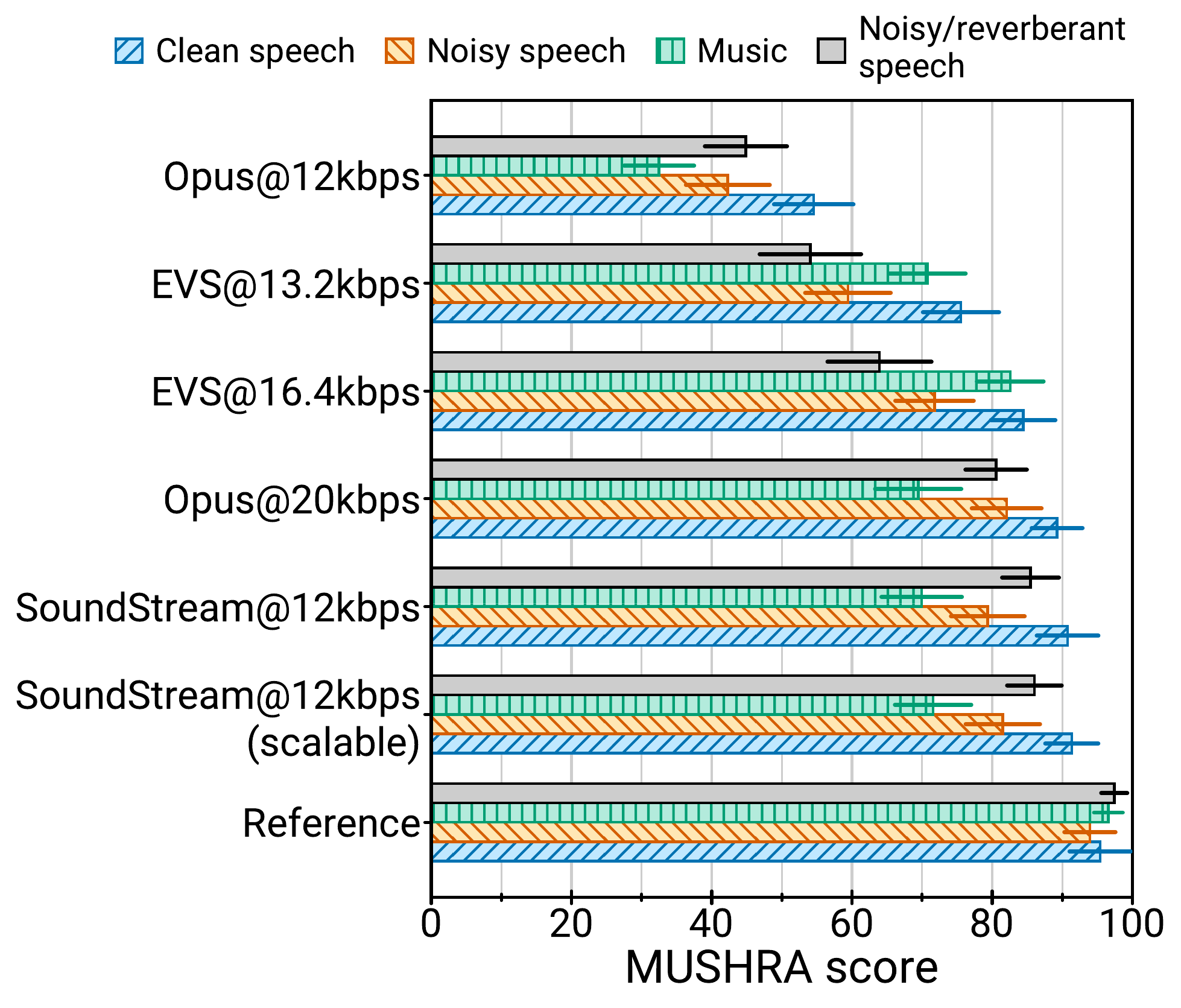}
        \caption{High bitrate.}
        \label{fig:mushra_12kbps_by_dataset}
    \end{subfigure}    
    \caption{Subjective evaluation results by content type. Error bars denote $95\%$ confidence intervals.}\label{fig:mushra_eval_by_dataset}
\end{figure*}

\subsection{Baselines}
Opus \cite{opus2012} is a versatile speech and audio codec supporting signal bandwidths from 4~kHz to 24~kHz and bitrates from 6~kbps to 510~kbps. 
Since its standardization by the IETF in 2012 it has  been widely deployed for speech communication over the internet. 
As the audio codec in applications such as Zoom and applications based on WebRTC ~\cite{WebRTC, rfc7478}, such as Microsoft Teams and Google Meet, Opus has hundreds
of millions of daily users. Opus is also one of the main audio codecs used in YouTube for streaming.
Enhanced Voice Services (EVS) \cite{evs2015} is the latest codec standardized by the 3GPP and was primarily designed for Voice over LTE (VoLTE).
Like Opus, it is a versatile codec operating at multiple signal bandwidths, 4~kHz to 20~kHz, and bitrates, 5.9~kbps to 128~kbps. It is replacing AMR-WB~\cite{bessette2002adaptive} and retains full backward operability. In this paper we utilize these two systems as baselines for comparison with the SoundStream codec.
For the lowest bitrates, we also compare the performance of the recently presented Lyra codec~\cite{kleijn2021lyra} which is an autoregressive generative codec operating at 3~kbps. We provide audio processed by \SoundStream and baselines at different bitrates on a public webpage\footnote{\url{https://google-research.github.io/seanet/soundstream/examples/}}.

\section{Results}
\label{sec:results}

\subsection{Comparison with other codecs}
Figure~\ref{fig:mushra_eval} reports the main result of the paper, where we compare \SoundStream to Opus and EVS at different bitrates. Namely, we repeated a subjective evaluation based on a MUSHRA-inspired crowdsourced scheme, when \SoundStream operates at three different bitrates: i) low ($3\,$kbps); ii) medium ($6\,$kbps); iii) high ($12\,$kbps). 
Figure~\ref{fig:mushra_3kbps} shows that \SoundStream at $3\,$kbps significantly outperforms both Opus at $6\,$kbps and EVS at $5.9\,$kbps (i.e., the lowest bitrates at which these codecs can operate), despite using half of the bitrate. To match the quality of \SoundStream, EVS needs to use at least $9.6\,$kbps and Opus at least $12\,$kbps, i.e., 3.2$\times$ to 4$\times$ more bits than \SoundStream. We also observe that \SoundStream outperforms Lyra when they both operate at $3\,$kbps.
We observe similar results when \SoundStream operates at $6\,$kbps and $12\,$kbps. At medium bitrates, EVS and Opus require, respectively, 2.2$\times$ to 2.6$\times$ more bits to match the same quality. At high bitrates, 1.3$\times$ to 1.6$\times$ more bits. 

Figure~\ref{fig:mushra_eval_by_dataset} illustrates the results of the subjective evaluation by content type. We observe that the quality of \SoundStream remains consistent when encoding clean speech and noisy speech. In addition, \SoundStream can encode music when using as little as $3\,$kbps, with quality significantly better than Opus at $12\,$kbps and EVS at 5.9 kbps. This is the first time that a codec is shown  to operate on diverse content types at such a low bitrate.

\subsection{Objective quality metrics}

\begin{figure*}[t]
    \centering
    \includegraphics[width=0.95\textwidth]{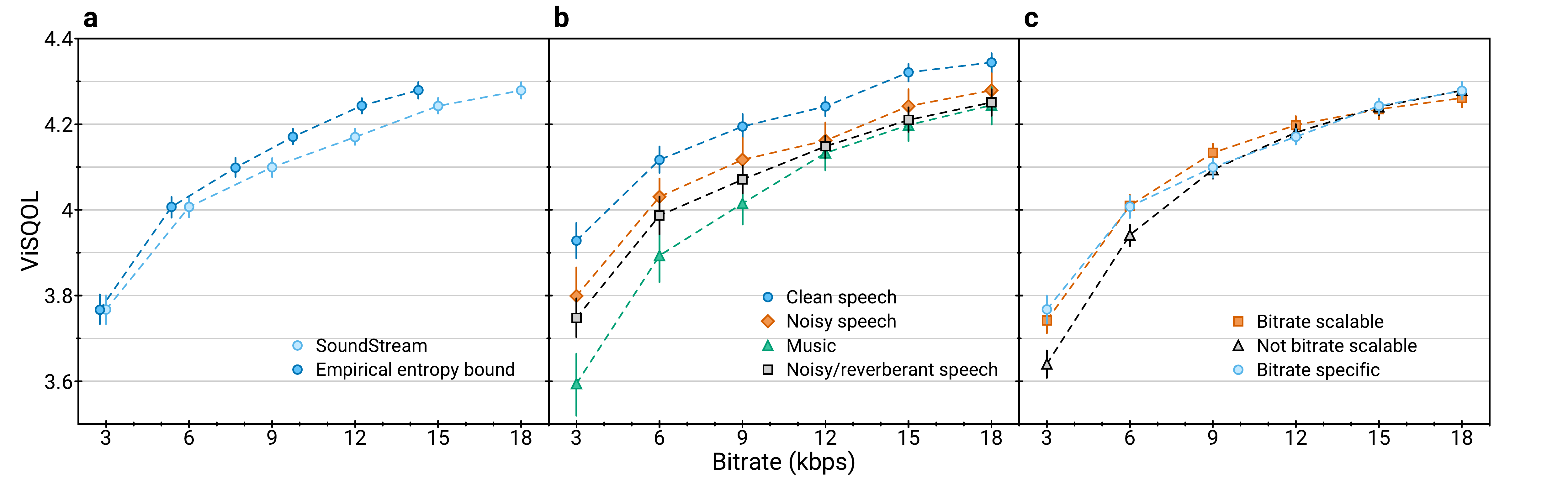}
    \caption{\visqol vs. bitrate. \textbf{a)} SoundStream performance on test data, comparing the actual bitrate with the potential bitrate savings achievable by entropy coding \textbf{b)} \visqol scores by content type \textbf{c)} Comparison of \SoundStream models that are trained at $18\,$kbps with quantizer dropout (bitrate scalable), without quantizer dropout (not bitrate scalable) and evaluated with a variable number of quantizers, or trained and evaluated at a fixed bitrate (bitrate specific). Error bars denote $95\%$ confidence intervals.}
    \label{fig:visqol_vs_rate}
\end{figure*}

Figure~\ref{fig:visqol_vs_rate}a shows the rate-quality curve of \SoundStream over a wide range of bitrates, from $3\,$kbps to $18\,$kbps. We observe that quality, as measured by means of \visqol, gracefully decreases as the bitrate is reduced and it remains above 3.7 even at the lowest bitrate. In our work, \SoundStream operates at constant bitrate, i.e., the same number of bits is allocated to each encoded frame. At the same time, we measure the bitrate lower bound by computing the empirical entropy of the quantization symbols of the vector quantizers, assuming each vector quantizer to be a discrete memoryless source, i.e., no statistical redundancy is exploited across different layers of the residual vector quantizer, nor across time. 
Figure~\ref{fig:visqol_vs_rate}a indicates a potential rate saving between 7\% and 20\%. 

We also investigate the rate-quality tradeoff achieved when encoding different content types, as illustrated in Figure~\ref{fig:visqol_vs_rate}b. Unsurprisingly, the highest quality is achieved when encoding clean speech. 
Music represents a more challenging case, due to its inherent diversity of content.

\subsection{Bitrate scalability}
\label{subsec:bitrate_scalability}

We investigate the bitrate scalability provided by training a single model that can serve different bitrates. To evaluate this aspect, for each bitrate $\bitrate$ we consider three \SoundStream configurations: a) a non-scalable model trained and evaluated at bitrate $\bitrate$ (\textit{bitrate specific}); b) a non-scalable model trained at 18\,kbps and evaluated at bitrate $\bitrate$ by using only the first $\numselectedquantizers$ quantizers during inference (\textit{18 kbps - no dropout}); c) a scalable model trained with quantizer dropout and evaluated at bitrate $\bitrate$ (\textit{bitrate scalable}). Figure \ref{fig:visqol_vs_rate}c shows the \visqol scores for these three scenarios. Remarkably, a model trained specifically at 18\,kbps retains good performance when evaluated at lower bitrates, even though the model was not trained in these conditions. Unsurprisingly, the quality drop increases as the bitrate decreases, i.e., when there is a more significant difference between training and inference. This gap vanishes when using the quantizer dropout strategy described in 
Section~\ref{subsec:vector_quantizer}. Surprisingly, the bitrate scalable model seems to marginally outperform bitrate specific models at 9\,kbps and 12\,kbps. This suggests that quantizer dropout, beyond providing bitrate scalability, may act as a regularizer. 

We confirm these results by including the bitrate scalable variant of \SoundStream in the MUSHRA subjective evaluation (see Figure \ref{fig:mushra_eval}). When operating at $3\,$kbps, the bitrate scalable variant of \SoundStream is only slightly worse than the bitrate specific variant. Conversely, both at 6~kbps and 12~kbps it matches the same quality as the bitrate specific variant. 

\subsection{Ablation studies}
We carried out several additional experiments to evaluate the impact of some of the design choices applied to \SoundStream. Unless stated otherwise, all these experiments operate at $6\,$kbps. 

\label{subsec:ablations}
\noindent\textbf{Advantage of learning the encoder}
-- We explored the impact of replacing the learnable encoder of \SoundStream with a fixed mel-filterbank, similarly to Lyra \cite{kleijn2021lyra}. We learned both the quantizer and the decoder and observed a significant drop in objective quality, with \visqol going from 3.96 to 3.33. Note that this is significantly worse than what can be achieved when learning the encoder and halving the bitrate (i.e., \visqol equal to 3.76 at $3\,$kbps). 
This demonstrates that the additional complexity of having a learnable encoder translates to a very significant improvement in the rate-quality trade-off.

\noindent\textbf{Encoder and decoder capacity} --
\begin{table}[t]
    \centering
    \caption{Audio quality (\visqol) and model complexity (number of parameters and real-time factor) for different capacity trade-offs between encoder and decoder, at 6kbps.}
    \begin{tabular}{c|c|c|c|c|c}
    \hline\hline
        $\baseconvdepth_{\enc}$ & $\baseconvdepth_{\dec}$ & \#Params & RTF (enc) & RTF (dec) &ViSQOL \\
    \hline\hline
        32 & 32 & \SI{8.4}{\mega\params} & 2.4$\times$ & 2.3$\times$ & 4.01 $\pm$ 0.03\\
        16 & 16 & \SI{2.4}{\mega\params} & 7.5$\times$ & 7.1$\times$ & 3.98 $\pm$ 0.03\\
        \hline\hline
        \multicolumn{4}{c}{Smaller encoder} \\ \hline
        16 & 32 & \SI{5.5}{\mega\params} & 7.5$\times$ & 2.3$\times$ & 4.02 $\pm$ 0.03\\
        8 & 32 & \SI{4.8}{\mega\params} & 18.6$\times$ & 2.3$\times$ & 3.99 $\pm$ 0.03\\
        \hline\hline
         \multicolumn{4}{c}{Smaller decoder} \\ \hline
        32 & 16 & \SI{5.3}{\mega\params} & 2.4$\times$ & 7.1$\times$ & 3.97 $\pm$ 0.03\\
        32 & 8 & \SI{4.4}{\mega\params} & 2.4$\times$ &  17.1$\times$ & 3.90 $\pm$ 0.03\\
    \hline\hline
    \end{tabular}
    \label{tab:capacity}
\end{table}
The main drawback of using a learnable encoder is the computational cost of the neural architecture, which can be significantly higher than computing fixed, non-learnable features such as mel-filterbanks. For \SoundStream to be competitive with traditional codecs, not only should it provide a better perceptual quality at an equivalent bitrate, but it must also run in real-time on resource-limited hardware. Table \ref{tab:capacity} shows how computational efficiency and audio quality are impacted by the number of channels in the encoder $\baseconvdepth_{\enc}$ and the decoder $\baseconvdepth_{\dec}$. We measured the real-time factor (RTF), defined as the ratio between the temporal length of the input audio and the time needed for encoding/decoding it with \SoundStream. We profiled these models on a single CPU thread of a Pixel4 smartphone. We observe that the default model ($\baseconvdepth_{\enc} = \baseconvdepth_{\dec} = 32$) runs in real-time (RTF $> 2.3\times$). Decreasing the model capacity by setting $\baseconvdepth_{\enc} = \baseconvdepth_{\dec} = 16$ only marginally affects the reconstruction quality while increasing the real-time factor significantly (RTF $> 7.1\times$).
We also investigated configurations with asymmetric model capacities. Using a smaller encoder, it is possible to achieve a significant speedup without sacrificing quality (\visqol drops from 3.96 to 3.94, while the encoder RTF increases to $18.6 \times$). Instead, decreasing the capacity of the decoder has a more significant impact on quality (\visqol drops from 3.96 to 3.84). This is aligned with recent findings in the field of neural image compression~\cite{mentzer2020highfidelity}, which also adopt a lighter encoder and a heavier decoder.

\begin{figure*}[t]
    \centering
    \includegraphics[width=0.95\textwidth]{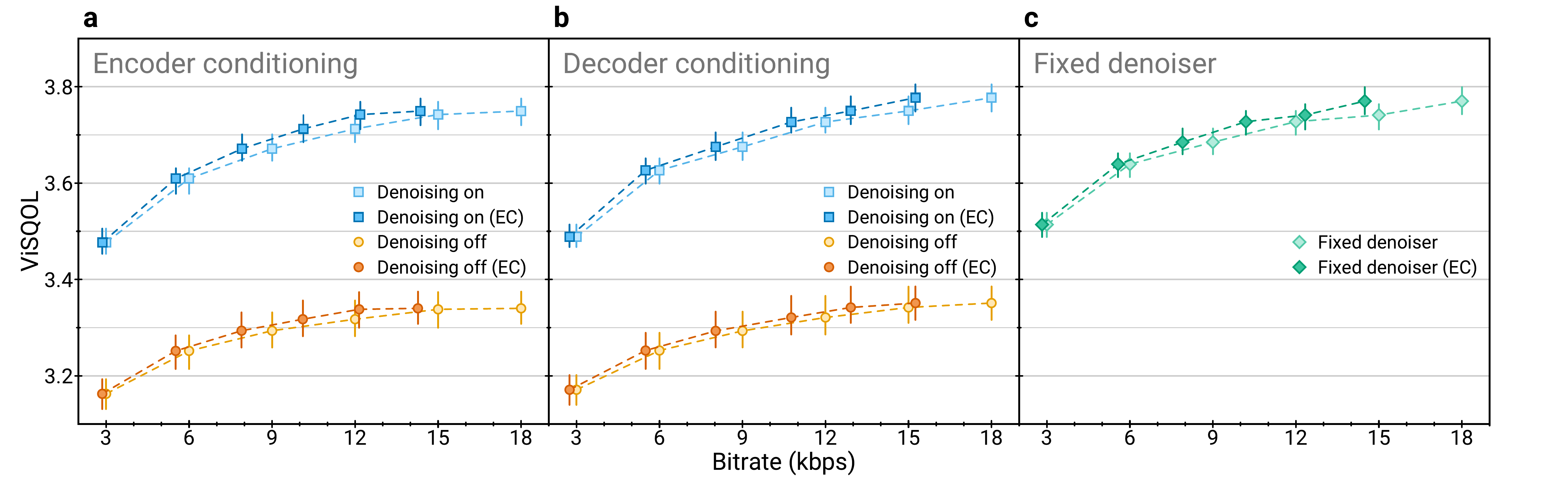}
    \caption{Performance of \SoundStream when performing joint compression and background noise suppression, measured by ViSQOL scores at different bitrates. We compare three variants: \textbf{a)} flexible denoising, where the conditioning is added at the encoder side; \textbf{b)} flexible denoising, where the conditioning is added at the decoder side; and \textbf{c)} fixed denoising, where the model was trained to always produce clean outputs. For all models we also  report the potential bitrate savings achievable by entropy coding (EC). Error bars denote $95\%$ confidence intervals.
    }
    \label{fig:soundstream_denoising_performance}
\end{figure*}

\noindent\textbf{Vector quantizer depth and codebook size} --
The number of bits used to encode a single frame is equal to $\numquantizers \log_{2}\codebooksize$, where $\numquantizers$ denotes the number of quantizers and $\codebooksize$ the codebook size. Hence, it is possible to achieve the same target bitrate for different combinations of $\numquantizers$ and $\codebooksize$. 
Table~\ref{tab:num_quantizers} shows three configurations, all operating at $6\,$kbps. As expected, using fewer vector quantizers, each with a larger codebook, achieves the highest coding efficiency at the cost of higher computational complexity. Remarkably, using a sequence of 80 1-bit quantizers leads only to a modest quality degradation. This demonstrates that it is possible to successfully train very deep residual vector quantizers without facing optimization issues. On the other side, as discussed in Section~\ref{subsec:vector_quantizer}, growing the codebook size can quickly lead to unmanageable memory requirements. Thus, the proposed residual vector quantizer offers a practical and effective solution for learning neural codecs operating at high bitrates, as it scales gracefully when using many quantizers, each with a smaller codebook.

\begin{table}[t]
    \centering
    \caption{Trade-off between residual vector quantizer depth and codebook size at $6\,$kbps.}
    \begin{tabular}{l|c|c|c}
    \hline\hline
    Number of quantizers $\numquantizers$ & 8 & 16 & 80 \\
    Codebook size $\codebooksize$ & 1024 & 32 & 2 \\
    ViSQOL & 4.01 $\pm$ 0.03 & 3.98 $\pm$ 0.03 & 3.92 $\pm$ 0.03\\
    \hline\hline
    \end{tabular}
    \label{tab:num_quantizers}
\end{table}

\noindent\textbf{Latency} --
\begin{table}[t]
    \centering
    \caption{Audio quality (\visqol) and real-time factor for different levels of architectural latency, defined by the total striding factor of the encoder/decoder, at $6\,$kbps.}
    \begin{tabular}{c|c|c|c|c|c}
    \hline\hline
        Strides & Latency & $\numquantizers$ & RTF (enc) & RTF (dec) &ViSQOL \\
    \hline\hline
       $(1, 4, 5, 8)$ & 7.5\si{\milli\second} & 4 & 1.6$\times$ & 1.5$\times$ & 4.01 $\pm$ 0.02\\
       $(2, 4, 5, 8)$ & 13\si{\milli\second} & 8 & 2.4$\times$ & 2.3$\times$ & 4.01 $\pm$ 0.03\\
       $(4, 4, 5, 8)$ & 26\si{\milli\second} & 16 & 4.1$\times$ & 4.0$\times$ & 4.01 $\pm$ 0.03\\
    \hline\hline
    \end{tabular}
    \label{tab:latency}
\end{table}
The architectural latency $\ratio$ of the model is defined by the product of the strides, as explained in Section \ref{subsec:encoder}. In our default configuration, $\ratio=2\cdot 4 \cdot 5 \cdot 8 = 320$ samples, which means that one frame corresponds to 13.3\si{\milli\second} of audio at $24\,$kHz. The bit budget allocated to the residual vector quantizer needs to be adjusted based on the target architectural latency. For example, when operating at $6\,$kbps, the residual vector quantizer has a budget of 80 bits per frame. If we double the latency, one frame corresponds to 26.6\si{\milli\second}, so the per-frame budget needs to be increased to 160 bits. Table~\ref{tab:latency} compares three configurations, all operating at $6\,$kbps, where the budget is adjusted by changing the number of quantizers, while keeping the codebook size fixed. We observe that these three configurations are equivalent in terms of audio quality. At the same time, increasing the latency of the model significantly increases the real-time factor, as encoding/decoding of a single frame corresponds to a longer audio sample. 

\subsection{Joint compression and enhancement}
\label{subsec:joint_compression_enhancement}
We evaluate a variant of \SoundStream that is able to jointly perform  compression and background noise suppression, which was trained as described in Section~\ref{subsec:conditioning}. We consider two configurations, in which the conditioning signal is applied to the embeddings: i) one where the conditioning signal is added at the encoder side, just before quantization; ii) another where it is added at the decoder side. For each configuration, we train models at different bitrates. For evaluation we use $1000$~samples of noisy speech, generated as described in Section~\ref{subsec:datasets} and compute ViSQOL scores when denoising is enabled or disabled, using clean speech references as targets. Figures~\ref{fig:soundstream_denoising_performance} shows a substantial improvement of quality when denoising is enabled, with no significant difference between denoising either at the encoder or at the decoder. %
We observe that the proposed model, which is able to flexibly enable or disable denoising at inference time, does not incur a cost in performance, when compared with a model in which denoising is always enabled. This can be seen comparing Figure~\ref{fig:soundstream_denoising_performance}c with Figure~\ref{fig:soundstream_denoising_performance}a and Figure~\ref{fig:soundstream_denoising_performance}b.

We also investigate whether denoising affects the potential bitrate savings that would be achievable by entropy coding. To evaluate this aspect, we first measured the empirical probability distributions $p_i^{(q)}, \, i = 1\ldots \codebooksize, \, q = 1\ldots \numquantizers$ on 3200 samples of training data. Then, we measured the empirical distribution $r_i^{(q)}$ on the $1000$~test samples and computed the cross-entropy $H(r, p) = -\sum_{i,q} r_i^{(q)} \log_2 p_i^{(q)}$, as an estimate of the bitrate lower bound needed to encode the test samples. Figure~\ref{fig:soundstream_denoising_performance} shows that both the encoder-side denoising and fixed denoising offer substantial bitrate savings when compared with decoder-side denoising. Hence, applying denoising before quantization leads to a representation that can be encoded with fewer bits. 

\subsection{Joint vs. disjoint compression and enhancement}
\label{subsec:joint_vs_disjoint}

We compare the proposed model, which is able to perform joint compression and enhancement, with a configuration in which compression is performed by \SoundStream (with denoising disabled) and enhancement by a dedicated denoising model. For the latter, we adopt SEANet~\cite{tagliasacchi2020seanet}, which features a very similar model architecture, with the notable exception of skip connections between encoder and decoder layers and the absence of quantization. We consider two variants: i) one in which compression is followed by denoising (i.e., denoising is applied at the decoder side); ii) another one in which denoising is followed by compression (i.e., denoising is applied at the encoder side).

We evaluate the different models using the VCTK dataset~\cite{yamagishi2019vctk}, which was neither used for training \SoundStream nor SEANet. The input samples are $2\,$s clips of noisy speech cropped to reduce periods of silence and resampled at $24\,$kHz. For each of the four input signal-to-noise ratios ($0\,$dB, $5\,$dB, $10\,$dB and $15\,$dB), we run inference on $1000$~samples and compute \visqol scores. As shown in Table~\ref{tab:SoundStreamVsSEANet}, one single model trained for joint compression and enhancement achieves a level of quality that is almost on par with using two disjoint models. Also, the former requires only half of the computational cost and incurs no additional architectural latency, which would be introduced when stacking disjoint models. 
We also observe that the performance gap decreases as the input SNR increases.

\begin{table}[t]
\caption{Comparison of \SoundStream as a joint denoiser and codec with SEANet as a denoiser compressed by a \SoundStream codec at different signal-to-noise ratios. Uncertainties denote $95\%$ confidence intervals.}
\label{tab:SoundStreamVsSEANet}
\renewcommand{\arraystretch}{1.2}
\centering
\begin{tabular}{ c | c  c  c }
\hline
&  \multicolumn{3}{c}{ \visqol }\\
\hline
\multirow{2}{*}{Input SNR} & \multirow{2}{*}{\SoundStream} & \multirow{2}{*}{\shortstack{\SoundStream \\ $\to$SEANet}} & \multirow{2}{*}{\shortstack{SEANet \\ $\to$\SoundStream}} \\ \\
\hline
$0\,$dB & \confintsym{2.93}{0.02} & \confintsym{3.02}{0.03}  & \confintsym{3.05}{0.02} \\
$5\,$dB & \confintsym{3.18}{0.02} & \confintsym{3.30}{0.02}  & \confintsym{3.31}{0.02} \\
$10\,$dB & \confintsym{3.42}{0.02} & \confintsym{3.51}{0.02} & \confintsym{3.50}{0.02} \\
$15\,$dB & \confintsym{3.58}{0.02}  & \confintsym{3.64}{0.02}  & \confintsym{3.63}{0.02} \\
\hline
\end{tabular}
\end{table}

\section{Conclusions}
We propose \SoundStream, a novel neural audio codec that outperforms state-of-the-art audio codecs over a wide range of bitrates and content types. \SoundStream consists of an encoder, a residual vector quantizer and a decoder, which are trained end-to-end using a mix of adversarial and reconstruction losses to achieve superior audio quality. The model supports streamable inference and can run in real-time on a single smartphone CPU. When trained with quantizer dropout, a single \SoundStream model achieves bitrate scalability with a minimal loss in performance when compared with bitrate-specific models. In addition, we show that it is possible to combine compression and enhancement in a single model without introducing additional latency.

\section*{Acknowledgments}
The authors thank Yunpeng Li, Dominik Roblek, Félix de Chaumont Quitry and Dick Lyon for their feedback on this work.

\balance

\bibliography{main}
\bibliographystyle{IEEEtran}

\end{document}